\documentclass[11pt,twoside]{article}

\usepackage{asp2006}

\begin{document}
\title{The Abundances of Oxygen and Carbon in the Solar Photosphere}   
\author{Carlos Allende Prieto}   
\affil{The University of Texas, Austin, Texas 78712}    

\begin{abstract}

A series of recent studies has placed the best estimates of the photospheric
abundances of carbon and oxygen at $\log \epsilon =8.39$ and $8.66$,
respectively. These values are $\sim 40$ \% lower than earlier 
estimates. A coalition of corrections 
due to the adoption of an improved model atmosphere, 
updated atomic data and non-LTE corrections, and a reevaluation of 
the effect of blending features, is responsible for the change. 
The adopted hydrodynamical model of the 
solar surface is an important element to the update, 
but using a theoretical 1D  model
atmosphere leads  to an average oxygen abundance modestly 
increased by 0.09 dex, and a carbon abundance only 0.02 dex higher.
Considering a state-of-the-art 3D hydrodynamical model of the
solar surface yields consistent results 
from different sets of atomic and molecular lines. Systematic errors
are likely to dominate the final uncertainties, 
but the available information indicates they are limited to $<0.1$ dex.
The new abundances are closer to expectations based on the compositions 
of other nearby objects, although a fully consistent picture, considering 
galactic chemical evolution and diffusion at the bottom of the solar 
convection zone, is still lacking.


\end{abstract}


\section{Introduction}

Spectroscopic analyses of normal hydrogen-rich 
stars naturally derive abundance ratios to hydrogen, as this element 
(either in atoms or forming H$^{-}$) dominates the opacity 
at optical and longer wavelengths.
An abundance determination is based upon a physical model of the
outer layers of the star, from where the observed photons escape.
Usually, such model relies on the assumption of plane-parallel 
homogeneous layers in hydrostatic and thermodynamic balance. 
The complexities of a real star's envelope (e.g. hydrodynamics,  
magnetic fields, and rotation) are neglected, in order to make
the problem tractable, and the model has only a few parameters 
such as the total energy flux and surface gravity, which can be
determined from observations.

A model is constructed assuming an initial set of chemical abundances.
The negative temperature gradient 
in the stellar atmosphere creates absorption lines 
which strength is proportional
to the absorber's density relative to hydrogen.
The actual abundance determination is performed by calculating a spectrum 
based on the adopted model atmosphere, comparing it to the observed lines, 
and adjusting the abundance as needed.
If necessary, the cycle is iterated by computing a new model, 
until convergence. 
Besides the model, transition probabilities are 
the most critical ingredient, 
and despite the success of quantum mechanics, accurate
calculations are still only possible for relatively simple atoms and
molecules. For others, we rely on laboratory measurements.

Systematic and random errors in the observations
have steadily been reduced in parallel with progress in instrumentation. 
Errors associated with the underlying approximations in the interpretation
of spectra are a more hairy issue. Grevesse \& Sauval (1999) examined
the history of solar iron abundance determinations starting with the 
analysis by Henry Norris Russell in 1929. 
The most salient feature was a fast increase of about an order
of magnitude between the results derived in the 60's and those
published in the 80's, which was caused by erroneous transition probabilities.
By looking at their plot, one would suggest that analyses may be converging at 
$\log \epsilon$\footnote{
$\epsilon {\rm (X)} = \frac{\rm N(X)}{\rm N(H)} \times 10^{12}$, 
where N represents number density.
}(Fe) $\simeq 7.5$,
with $\sigma \simeq 0.1$ dex. 

More recently, Gustafsson (2004) has made a similar study for the case of
oxygen. The situation with this element is somewhat different, and could be 
described as a steady decrease with significant scatter. 
Unlike the iron case, 
oxygen is a simpler atom which is successfully modeled
under LS coupling, and Gustafsson attributes the scatter mainly to 
differences in the analysis: employed spectral lines, model atmospheres, and 
the calculation of theoretical spectra. 

Most
elements observed and studied in the solar photosphere have a 
similar abundances pattern as those found in a special type of meteorites: 
the CI (type I, C1, or Ivuna-type) carbonaceous chondrites 
(see, e.g., Anders \& Grevesse 1989). 
This very rare class of meteorites, of which only five falls are known, 
appears to have preserved the original composition
of the solar nebula, with the exception of the most volatile elements:
H, C, N, O, and the noble gases. Meteoritic abundances are usually given 
relative to silicon, and using this element to bridge with photospheric
abundances, the agreement is better than 10\% for 31 elements 
and better then 15\% for 41 (Lodders 2003). 
The number density ratio Fe/Si derived in recent analyses of the solar
spectrum  is 0.87, while in CI meteorites is 0.863 
(Asplund et al. 2000b, Asplund 2000, Lodders 2003) -- a stunning agreement.

Unfortunately, the carbon and oxygen cases have no {\it solid} reference, 
and we are left to figure out their abundances from 
solar observations alone.
We will not discuss whether the larger scatter in 
the historical evolution of the solar oxygen abundance determinations
is related to this fact, and we will focus on the technical 
aspects. The motivation for this review is, primarily, 
the  recent jumps
in the value of the solar carbon and oxygen abundances as estimated from 
photospheric oxygen lines in the solar spectrum. 
The recent series of studies on this subject 
was driven by the availability of three-dimensional
radiation hydrodynamical simulations of solar surface convection
(for a review see Asplund 2005).
As we shall see, however, the new models are only partly responsible
for the proposed abundance change. We will discuss how confident we can
be about the new values, and what pieces of the puzzle still do not fit,
trying to put the finger in the wound in order 
to stimulate further work in areas where needed the most.

The first section is devoted to the difficulties
involved in an analysis of oxygen lines in the solar spectrum. Section
3 repeats the same exercise for carbon.
Section 4 is an attempt to summarize the main uncertainties still
present in current abundance 
determinations, and suggest directions for future
research. Section 5 looks at other objects in the solar neighborhood
to see whether current values of the solar abundance make sense
in a broader context. Section 6 closes this piece with a discussion 
and a summary.

\section{Spectroscopic determination of the solar photospheric \\
oxygen abundance}

Despite an atomic system amenable to accurate
quantum mechanical calculations, the analysis of oxygen in the solar
spectrum involves extra difficulties compared to the case of iron.
The solar spectrum offers only a handful
of oxygen absorption lines, and they tend to form far from 
local thermodynamic equilibrium (LTE) conditions, which makes their 
calculation more difficult and more dependent on collisional 
interactions.
Notable exceptions are the forbidden transitions
that connect the atom's lowest states. 

\subsection{Atomic forbidden lines}

Of the six forbidden transitions that connect the lowest three
states of oxygen, there are three that fall in the optical and
(albeit weak) are measurable in the solar spectrum (Lambert 1978).
The $^1$D to $^1$S transition at 557.7 nm 
is badly blended with C$_2$ lines. The other two lines 
($^3$P J=2,1 to $^1$D at  630.0 nm  and 636.3 nm) 
coincide in wavelength with a Ni I and a  couple of 
CN transitions, respectively, but the contribution of these blends, 
although uncertain, is less significant.  After estimating the strength
of the potential blending features, Lambert (1978)
used these weak lines to derive an oxygen abundance of 
$\log \epsilon$(O)$= 8.92$. 

The recent turmoil started precisely with a reanalysis of  
the forbidden line at 630.0 nm. Despite the transition probability 
of the Ni I remained highly uncertain, Allende Prieto, Lambert \&
Asplund (2001) realized that its strength could be inferred,
independently of the oxygen abundance, by taking advantage of 
the new generation of hydrodynamical models of the solar surface.
Solar photospheric absorption lines are asymmetric and blue-shifted
from their rest wavelength by up to 500 m s$^{-1}$ 
as a result of convective motions in the
solar envelope -- what is observed in images of the solar surface
as granulation. Classical model atmospheres, built under the assumption
of hydrostatic equilibrium, predict perfectly symmetric line profiles.
State-of-the-art hydrodynamical models, in turn, predict 
convective line asymmetries and shifts in good agreement with those
observed, typically within $50-100$ m s$^{-1}$ 
(Allende Prieto et al. 2002; Asplund et al. 2000a).

The rest wavelengths of the oxygen and nickel transitions were
already known with an accuracy of $0.2-0.3$ pm 
(1 pm $\equiv 10^{-12}$ m).
By modeling the oxygen abundance, the strength of the Ni I blending
contribution ($gf \times \epsilon$(Ni)), and a continuum normalization
factor for the observed line profile, Allende Prieto et al. 
concluded that evidence pointed to a lower 
oxygen abundance than earlier estimates, 
$\log \epsilon$(O)$=8.69 \pm 0.05$ dex.
The revision was associated with a much larger contribution 
from the Ni I line ($\sim 30$ \%, or an abundance change of 0.13 dex), 
and a different thermodynamical
structure of the hydrodynamical simulations compared to the
classical model atmospheres used previously (0.08 dex), both working in the
same direction.

More recently, Johansson et al. (2003) performed laboratory 
measurements of the Ni I transition overlapped with [OI] 630.0 nm,
providing a transition probability as well as wavelengths
for the components  associated with the two most abundant isotopes, 
$^{58}$Ni and $^{60}$Ni, which 
together account for  over 94 \% of the nickel on Earth.
If the analysis of the solar feature is repeated accounting
for the isotopic shift ($\delta\lambda \simeq$ 2 pm), 
the result remains the same. However, if 
the photospheric nickel abundance adopted by 
Grevesse \& Sauval (1998) is combined with the line strength inferred 
by Allende Prieto et al., the resulting $f-$value would be $\sim$ 0.2 dex
smaller than the laboratory measurement, which has an uncertainty of
0.05 dex. Embracing the laboratory $f-$value and the photospheric
nickel abundance would imply a further
reduction to the oxygen abundance by about 18 \% or 0.08 dex, but
Koesterke et al. (see poster in these proceedings) 
have found that the resulting spectrum is not 
compatible with the observed feature. 
It remains to be seen 
if a 3D-based analysis of nickel in the Sun reduces this problem,
but if we compare the meteoritic Ni/Si ratio of 0.048 with the
photospheric value of $\simeq 0.05-0.06$, there is not enough room
to accommodate the change in strength implied by the updated value
of the  transition probability.

A subsequent
reanalysis of the weaker 636.3 nm forbidden transition,
using the same model
yielded an abundance consistent with the value from the transition
at 630.0 nm (Asplund et al. 2004). We recall that this
feature was blended with two CN transitions, and the correction in 
this case is mainly linked to the change in model atmosphere, i.e.
1D vs. 3D, given the large sensitivity of molecular species  
to temperature inhomogeneities. 
The hydrodynamical simulations 
have a horizontal average temperature\footnote{
Horizontal average, in this context, refers to the mean value  
over  a  surface with a constant optical depth, rather than over 
a constant geometrical depth.
} lower than classical one-dimensional models in the outer layers, 
and temperature inhomogeneities  further enhance the strength
of the CN blending transition, resulting on a lower oxygen abundance.

\subsection{Atomic permitted lines}

Asplund et al. (2004) collated the list of oxygen lines observed 
in the solar spectrum and selected a subset of six permitted
transitions with accurate $f-$values and clean profiles. 
Some of these lines are
suspected to suffer marked departures from local thermodynamical
equilibrium since the early work by Mihalas \& Henshaw (1966)
and Altrock (1968).
The analysis of oxygen lines in a three-dimensional solar-like 
model atmosphere considering 
non-LTE was pioneered by Kiselman \& Nordlund (1995).
Interestingly, the calculated non-LTE corrections for these lines
from classical 1D models and from the hydrodynamical simulations 
are quite similar, and
always work reducing the abundances derived assuming 
LTE by an amount that ranges between 0.03 and 0.3 dex.
Purely 3D effects are usually smaller ($< 0.03$ dex).

Approximate cross-sections for the collisional processes are
typically used in the non-LTE calculations. The relevant
particles are electrons and hydrogen atoms. Measurements of
electron collisional excitation rates are limited to 
transitions from the ground state. 
Widely-used general recipes based on the van Regemorter (1962) formula 
can be widely applied, but are uncertain by a few orders of magnitude
(see, e.g., Allende Prieto et al. 2003). 
More refined calculations  exist for some transitions.
Recent independent calculations for low-lying states 
by Zatsarinny \& Tayal (2003)  and Barklem (2007) typically agree 
within 70 \%. 

The situation is worse for hydrogen collisions.
A recipe originally proposed by Steenbock \& Holweger (1984) based 
on Drawin's approximation (Drawin 1968, 1969) has been broadly used, 
despite it has been shown to 
overestimate the rates by several orders of magnitude for alkalis 
(Belyaev \& Barklem 2003). This situation has led some modelers to neglect
hydrogen collisions altogether in their calculations. While their effect
is extremely important for low metallicity stars, it induces only 
modest changes in the inferred oxygen populations under solar conditions.

The non-LTE abundances derived by Asplund et al. are in the range 
8.60--8.65, depending on the line, with an average result of
8.64 ($\sigma = 0.02$) dex. 
These calculations neglect
inelastic hydrogen collisions. The center-to-limb variation of
permitted oxygen lines has been repeatedly used to demonstrate 
the need to consider departures from LTE in their formation, 
but the same observations
also favor including hydrogen collisions 
(Allende Prieto, Asplund \& Fabiani Bendicho 2004).
If this path is taken, the average oxygen abundance from permitted
lines would increase to 8.70 ($\sigma = 0.04$) dex. 

As an example of the typical scatter found in oxygen non-LTE calculations 
in the literature, we examine the 777.4 nm line, 
part of the infrared ($^5$S$^{\rm o}$--$^5$P) triplet.
Asplund et al. inferred an abundance  correction of $-0.24$ dex for the
3D simulation, $-0.27$ dex for 
the Holweger-M\"uller model (Holweger \& M\"uller 1974), 
and $-0.23$ for a solar MARCS model; 
Allende Prieto et al. (2004), using a different model atom and statistical
equilibrium solver, derived a correction  of $-0.22$ dex for a 
 Kurucz solar model\footnote{\tt http://kurucz.harvard.edu/}. 
 These figures neglect inelastic hydrogen collisions.
Allende Prieto et al. found an abundance correction of $-0.16$ when
H collisions were included using the recipe mentioned above.
Another study by Holweger (2001), based on a third choice of model
atom and the Holweger-M\"uller atmosphere, but considering 
hydrogen collisions as above, arrived at a smaller correction 
of $-0.06$ dex. Hydrogen collisions in the non-LTE calculations 
are likely the dominant contribution that sets the systematic errors 
in the abundances from permitted lines at about $0.10$ dex.

\subsection{Molecular lines: OH}

Asplund et al. (2004) considered in their analysis OH pure rotation and OH 
vibration-rotation lines. The vast number of transitions and relatively
small overlapping problems in the infrared are an advantage compared to
using atomic lines. Sadly, molecule formation is extremely temperature
sensitive, and using a hydrodynamical model produces much larger
corrections than for atomic transitions.
The high sensitivity to temperature also implies a higher susceptibility
to the details involved in the numerical simulations, such as 
resolution or boundary conditions. The formation of the lines in 
higher layers and therefore lower density also increases the
risk for errors in the computed electron density, based on LTE and
chemical equilibrium for all relevant species. 

Despite the concerns,
the average abundance derived from the solar 3D simulation by
Asplund et al. from vibro-rotational and pure rotational transitions agree
remarkably well with each other: 8.61 ($\sigma=0.03$) and 
8.65 ($\sigma=0.02$), respectively. As expected, 
these figures involved substantial
corrections in comparison with a 1D analysis: they are lower 
by about 0.2 dex. 

The fact that the corrections associated with the introduction
of hydrodynamics are larger for molecular transitions 
than for atomic lines underlines that both indicators
cannot agree with both kinds of models. This is evident in 
Table 1. Indeed, the difference
between the oxygen abundance inferred from atomic and molecular 
transitions is significantly smaller with the new hydrodynamical model 
than with the classical semi-empirical Holweger-M\"uller model.
The solar 1D MARCS model studied by Asplund et al. turned out
to take a more moderate position: with this model, atomic O and
OH vibro-rotational lines converged
at an abundance of $8.72-8.74$, while only the OH rotational transitions
pointed to a higher value by 0.1 dex.

\begin{table}[!ht]
\caption{Average oxygen abundance (and standard deviation) 
inferred from different 
indicators. Adapted from Asplund et al. (2004).
The mean values in the last line of the table are a straight average
including the non-LTE calculations that neglect hydrogen collisions. 
There are no published calculations with the HM and MARCS
models including inelastic hydrogen collisions, but a study of the
infrared triplet with a solar Kurucz model and the 3D analysis for many lines
indicate that the resulting abundance 
will be $\simeq +0.05$ dex higher than neglecting
H collisions.}
\smallskip
\begin{center}
{\small
\begin{tabular}{cccc}
\tableline
\noalign{\smallskip}
lines 		&	3D	&	HM	&	MARCS \\
\noalign{\smallskip}
\tableline
\noalign{\smallskip}
[OI]		&	8.68 (0.01)&	8.76 (0.01)&	8.72 (0.01)\\
OI (without H coll.)&	8.64 (0.02)&	8.64 (0.08)&	8.72 (0.03)\\
OI (with H coll.)	&	8.70 (0.04)&	\dots	   &	\dots	   \\
OH vib-rot	&	8.61 (0.03)&	8.87 (0.03)&	8.74 (0.03)\\
OH rot		& 	8.65 (0.02)&	8.82 (0.01)&	8.83 (0.03)\\
\noalign{\smallskip}
\tableline
{\bf Average}	&	8.65 (0.03)&	8.77 (0.10)&    8.75 (0.05)\\
\noalign{\smallskip}
\tableline
\end{tabular}
}
\end{center}
\end{table}

\subsection{Molecular lines: CO}

CO transitions offer a less direct path to the oxygen and carbon 
abundances. In a very recent paper by Scott et al. (2006), 
a solar hydrodynamical
model, the same used in the studies described above,
was used to analyze $^{12}$C $^{16}$O,  $^{13}$C $^{16}$O, and 
$^{12}$C $^{18}$O vibration-rotation lines in the disk-center 
solar spectrum. These authors find that the strongest $^{12}$C $^{16}$O
lines, which form in very high layers, are not well reproduced by the model
when the same abundances derived from other atomic and molecular lines
are used, but weaker lines are perfectly consistent with the results
from other features. Not everybody is convinced, however, 
that a consistent picture is in place, and 
Ayres, Plymate \& Keller (2006) conclude that a semi-empirical
analysis of CO lines assuming the ratio C/O $=0.5$ results in an
oxygen abundance significantly higher than those derived by 
Asplund et al. (see also Ayres' contribution in this volume).

\section{Spectroscopic determination of the solar photospheric \\
carbon abundance}

The photospheric carbon abundance has also recently undergone a significant
revision from values in the range $\log \epsilon$(C)$= 8.5-8.6$ 
(Anders \& Grevesse 1989, Grevesse \& Sauval 1998, Holweger 2001) 
to $\log \epsilon$(C)$= 8.39$. The decrease was again proposed 
after a reanalysis of a forbidden line (872.7 nm), 
which in parallel to the case of oxygen,
has been shown to be fairly immune to departures from LTE (Allende Prieto,
Lambert \& Asplund 2002). No blending features were detected, but 
the revision was driven by improved quantum-mechanical
calculations of the transition probability ($0.07$ dex) and the
introduction of a hydrodynamical solar model ($0.08$ dex).

The analysis was later extended to include permitted lines, 
as well as CH vibro-rotational lines and C$_2$ electronic transitions
by Asplund et al. (2005).
Their study considered 16 atomic permitted transitions  in the optical
and infrared. For these lines, non-LTE abundance corrections
were estimated in the range $-0.03$ to $-0.15$ dex, depending on the line,
based on 1D models due to the need to
consider a complex model atom to properly account for the departures
from LTE. Here again, hydrogen collisions were not considered, and
if included with the Steenbock-Holweger recipe  
the non-LTE corrections  would likely be reduced by 
half (St\"urenburg \& Holweger 1990).

As for oxygen, the introduction of the hydrodynamical model
affects molecules the most, and a much more homogeneous answer 
is obtained with the 3D model than with the classical Holweger-M\"uller
model (see Table 2). With the 1D solar MARCS model, we find again
an intermediate level of scatter, which 
essentially leads to the same average abundance as in 3D.

\begin{table}[!ht]
\caption{Average carbon abundance (and standard deviation) 
inferred from different 
indicators. Data from Asplund et al. (2005).}
\smallskip
\begin{center}
{\small
\begin{tabular}{cccc}
\tableline
\noalign{\smallskip}
lines 		&	3D	&	HM	&	MARCS \\
\noalign{\smallskip}
\tableline
\noalign{\smallskip}
[CI]		&	8.39 (0.04)&	8.45 (0.04)&	8.40 (0.04)\\
CI (1D NLTE)	&	8.36 (0.03)&	8.39 (0.03)&	9.35 (0.03)\\	
CH vib-rot	&	8.38 (0.04)&	8.53 (0.04)&	8.42 (0.04)\\
C$_2$ Swan	&	8.44 (0.03)&	8.53 (0.03)&	8.46 (0.03)\\
\noalign{\smallskip}
\tableline
{\bf Average}		&	8.39 (0.03)&	8.48 (0.07)&	8.41 (0.05)\\
\noalign{\smallskip}
\tableline
\end{tabular}
}
\end{center}
\end{table}

\section{Where do we stand and what remains to be done.}

Most of the results discussed above come a set of recent papers
by the same group. Although several hydrodynamical simulations of
solar surface convection can be found in the literature,
only one of them has been extensively applied to the calculation
of detailed solar spectra. The exception to this statement is 
the independent study by  Holweger (2001; see also 
Steffen \& Holweger 2002) who analyzed several forbidden 
and permitted lines atomic lines considering 2D effects (in LTE), and 
1D non-LTE corrections to arrive at a mean oxygen abundance
of 8.74 (0.08) dex;  0.08 dex higher than the value recommended
by Asplund et al. (2004). 

Despite the remarkable success of the simulations of Stein \& Nordlund (1998)
and Asplund et al. in reproducing solar data and eliminating the need for
'ad hoc' turbulence, the lack of studies based on
alternative models is unsatisfactory. Hydrodynamics
are a big improvement, but the increased computational burden requires
sacrificing detail in the radiative term of the energy balance equation.
At least two of the most important tests available for validating a model of
the solar surface are yet to be done: center-to-limb variation and
absolute fluxes. The impact of the magnetic fields on the observed 
profiles is probably very small (Khomenko et al. 2005), but there is not yet 
consensus about the topology and strength of the magnetic fields
that permeate the solar photosphere.

Apart from a model of the lower atmosphere, many other important
ingredients are involved in the abundance determinations: partition
functions, excitation energies, damping constants, continuum opacities, 
etc. Most have non-linear effects on the inferred abundances, and a 
Monte-Carlo approach is necessary to derive reliable uncertainties.
As described in previous sections, blending features (especially 
relevant to weak oxygen forbidden lines), and departures from 
LTE (in particular, the impact of inelastic hydrogen collisions on
permitted lines) cannot yet be considered a solved problem.

A quick look at the straight average abundances in 
Tables 1 and 2 leads to an interesting conclusion. The proposed
abundances from a 3D analysis are only $\sim 0.1$ dex or less
lower than the equivalent results from a 1D analyses. The historical
account given in the previous sections shows that the predictive
power of a hydrodynamical model, and the high homogeneity 
of the abundances derived in 3D from different indicators 
have been instrumental in the revision.
Nonetheless, an important part of 
the responsibility  for the final outcome is linked to 
a combination of updated atomic data, recent high-quality 
observations, and non-LTE corrections.
As a result, if one were to choose ignoring the results based on
hydrodynamical models, sticking to the one-dimensional MARCS structures
\footnote{The Holweger-M\"uller model is not considered, as
it leads to a much larger scatter among different indicators.}, the 
preferred values for the oxygen and carbon abundances would still be 
$\log \epsilon$(O) $=8.75$, 
and 
$\log \epsilon$(O) $=8.41$, respectively.

\section{The solar neighborhood}
\label{neighbors}

We cannot seek help to determine or refine the solar abundances
from other nearby stars. But it is quite interesting to compare
the solar values with other objects in the solar neighborhood and
see if the overall abundances {\it make sense} in the context
of galactic chemical evolution. We briefly review results
from nearby  H II regions, the local ISM, massive stars, 
and point the reader to Jorge Melendez's poster (in these proceedings) 
for an analysis of disk giants.  A recent investigation of
solar analogs indicates that previously reported offsets between 
the abundance ratios in the Sun and  nearby  late-type
stars  are likely the result of systematic 
errors, and the Sun appears now chemically very similar to its 
G-type neighbors with the same iron abundance (Allende Prieto 2007).
A more detailed discussion about nearby FGK dwarfs is
deferred, given the difficulties to constrain ages from stellar
evolution models for unevolved stars.

\subsection{HII regions}

An extensive analysis of recombination transitions in
galactic HII regions has been recently performed by Esteban et al. (2005).
Unlike collisionally excited lines, the strength of recombination 
transitions is insensitive to the presence of 
suspected temperature fluctuations -- and therefore a more solid 
diagnostic. HII regions trace tightly the abundance gradient
of  carbon and oxygen in the galactic disk between 6.3 to 10.5 kpc from
the galactic center (d$\log \epsilon$/d$R = -0.10$ and $-0.04$ dex/kpc, 
respectively). Interpolating the results at the 
solar circle ($R \simeq 8$ kpc) carbon and oxygen abundances of 
8.57 and 8.69 dex are derived. 
Corrections of 0.08--0.10 dex are estimated for the the amounts 
trapped into dust.
The final figures are higher  than the preferred solar
photospheric abundances by 0.13 dex and 0.26 dex for carbon and oxygen,
respectively, matching  the increases due to chemical evolution 
predicted by Carigi et al. (2005) in the 4.6 Gyr since the formation of
the Sun (0.13 and 0.28 dex) but leave no room for diffusion at the bottom
of the convection zone expected by solar evolution models
(0.07 dex; see Lodders 2003).

\subsection{The local ISM}

The oxygen abundance in diffuse nearby ISM clouds has been surveyed
using weak absorption features in the UV. The results are
fairly uniform and yield an average oxygen abundance of 
$\log \epsilon$(O) $=8.54$ (Oliveira et al. 2005 and references therein). 
This value is
quite close to the revised photospheric abundance if $\sim 30$\%
of oxygen is depleted into dust grains, but leaves no room
for either diffusion in the Sun or chemical evolution in the ISM.
A larger depletion in dust grains in the form of silicates or
oxides seems unlikely (Linsky et al. 2006). Measuring the ratio
C/H is a harder task, but Slavin \& Frisch (2006) 
derived a lower limit of $\log \epsilon$(O) $=8.52$ 
and more likely $8.6-8.9$ for the local interstellar cloud 
toward $\epsilon$ CMa,
which could accommodate both a reduction in the solar photosphere and
an increase in the ISM since the solar formation.

\subsection{OB stars}

Massive stars can be used, similarly to H II regions, to trace the
{\it current} composition of the gas in the galactic disk at different 
galactocentric distances. Photospheric lines of carbon and oxygen
in these stars usually require substantial non-LTE corrections, which
were considered by 
Daflon  \& Cunha (2004) in a study sampling objects between 
$R= 5$ and 13 kpc from the galactic center. Although the scatter 
is significantly
larger than that found for H II regions, radial variations are clearly
detected with d$\log \epsilon$/d$R \sim -0.03$ dex/kpc for carbon and oxygen.
At the solar circle, the average abundances are about 
$0.1-0.2$ lower than solar photospheric values, which again,
taken at face value, cannot be reconciled with chemical
evolution in the ISM or metal diffusion at the bottom of the solar convection 
zone.

\section{Summary and conclusions}

Recent studies of the solar spectrum have arrived at
abundances of carbon and oxygen which are about 0.2 dex lower
than earlier estimates.
The revision comes as the result of a conspiracy of corrections, 
and it is the outcome of a comprehensive
and critical reanalysis of the available abundance indicators,
considering improved model atmospheres, updated atomic
data, non-LTE corrections (affecting  permitted atomic transitions), 
and a reevaluation of the effect of blending features. 
The presence of temperature
inhomogeneities has the largest impact on molecular species, while atomic
permitted lines are only weakly affected. Thus, 
the use of a hydrodynamical model of the solar
surface is an important element to the correction, 
but using a theoretical hydrostatic model
atmosphere leads  to an average oxygen abundance modestly 
increased by 0.09 dex, and a carbon abundance only 0.02 dex higher.

As oxygen and carbon are the most abundant elements after hydrogen
and helium, the revised abundances  have a significant impact on the
overall {\it metallicity} of the Sun. The metal mass fraction
is reduced from  the Grevesse \& Sauval (1998) reference of 
$Z= 0.017$  to 0.014 just due to the change in carbon and oxygen.
If other elements  recently revised are considered  then $Z$ 
will further decrease to 0.012 (Asplund et al. 2005). 

The revised values have been welcome by researchers on L and T
spectral types, as they lead to improvements in the comparison
with observed spectra (Burrows, Sudarsky \& Hubeny 2006). 
The new abundances are also more in line with  
those found in other objects in the solar neighborhood although, 
as discussed in \S 5, our picture of the
chemical patterns in the solar neighborhood is still incomplete;
overall, the solar neighborhood would be {\it more comfortable} with
an even lower  solar oxygen abundance.
On the other hand,  
when the updated abundances are used to calculate models 
of the solar interior,
the resulting structures are incompatible with helioseismic data 
(see, e.g., Bahcall et al. 2005,  Basu \& Antia 2004).
Models based on the new values predict the wrong value for the radius
of the convection zone, and the radial variation of the sound speed
inferred from seismic inversions is up to 10 \% different from 
model predictions, while the older abundances 
yielded a much better agreement with deviations of just a few percent.

If the revised abundances are correct, the blame must be somewhere
else. The reported problems could be ameliorated, if not cured, 
with a new source of extra opacity  below the convection zone.
A recent update from the Opacity Project goes in that direction,
but the calculations are still far from accounting for 
the necessary extra opacity (Seaton \& Badnell 2004, Badnell et al. 2005).
A significant increase of the solar neon abundance, which is 
inferred from the ratio Ne/O in solar
energetic particles and the photospheric oxygen abundance, has been
proposed (Antia \& Basu 2005, Bahcall, Basu, \& Serenelli 2005), 
and supported by analyses of nearby stars 
(Cunha et al. 2006, Drake \& Testa 2006; see also the corresponding
contributions in these proceedings), but this solution 
has found opposition from solar specialists (Schmelz, this volume; 
Young 2005) and may be incompatible with the
sound speed profile of the Sun (Delahaye \& Pinsonneault 2006).
Another possibility would be an enhancement of
opacity induced by gravity waves at the base of the convection
zone (Arnett, Meakin \& Young 2005, Press \& Rybicki 1981).
Most likely not all bets are on the table, but one hopes that
at the resolution of this puzzle something new will be learnt
about the Sun.

\acknowledgements 

It has been a pleasure to participate in this work 
with a number of colleagues and good friends, including 
Paul Barklem, Bengt Gustafsson, 
Katia Cunha, Pe\~na Fabiani, Ram\'on Garc\'{\i}a L\'opez, Nicolas
Grevesse, Dan Kiselman, Lars Koesterke, {\AA}ke Nordlund, Ivan Ram\'{\i}rez, 
Regner Trampedach, and especially   
Martin Asplund and David Lambert. I am also indebted to Robert Kurucz
and Roger Bell, as it was their estimate of the 
$f-$value of the Ni I blend what got this started. NSF 
(AST-0086321) and NASA continuing
support (NAG5-13057 and NAG5-13147) is greatly appreciated.


\end{document}